# Generator Contingency Modeling in Electric Energy Markets: Derivation of Prices via Duality Theory

N. G. Singhal, J. Kwon, and K. W. Hedman, *Member, IEEE*

*Abstract*-- Traditional electric energy markets do not explicitly model generator contingencies. To improve the representation of resources and to enhance the modeling of uncertainty, existing markets are moving in the direction of including generator contingencies and remedial action schemes within market auction models explicitly. This research contributes to the market design realm by providing detailed analysis of impending changes, it provides insightful guidance in understanding the market implications, and it provides recommendations on necessary changes to ensure a fair and transparent market structure. A primal (and the corresponding dual) formulation that accounts for the proposed changes to the auction model is provided to enable a theoretical analysis of the anticipated changes including the effect on market prices, settlements, and revenues. The derivation of the prices and the dual formulation are based on leveraging duality theory from linear optimization theory. A comparison to existing market structures is also included. The primary impact of the proposed changes includes the addition of a new congestion component within the traditional locational marginal price, which reflects the influence of congestion during the post-contingency states for the modeled critical generator contingencies.

*Index Terms*-- Contingency reserve, generator contingency modeling, power system reliability, electricity market design.

## I. Nomenclature

*Sets and Indices*

| | |
|---|---|
| $c$ | Contingency index. |
| $C^g$ | Set of generator contingencies. |
| $C^{g^{crt}}$ | Subset of critical (credible) generator contingencies, $C^{g^{crt}} \subseteq C^g$. |
| $k$ | Transmission asset (line or transformer) index. |
| $K$ | Set of transmission assets. |
| $K^{crt}$ | Subset of critical transmission assets, $K^{crt} \subseteq K$. |
| $n$ | Node (or bus) index. |
| $N$ | Set of nodes. |
| $n'(c)$ | Node index for generator loss under contingency $c$. |
| $S^{FR}$ | Set of nodes that have generators with frequency response capability. |

*Parameters*

| | |
|---|---|
| $c_n$ | Operating cost for the generator at node $n$. |
| $\overline{D_n}$ | Fixed real power demand at node $n$. |
| $GDF_{n'(c),n}$ | Generation loss distribution factor for contingency $c$ at node $n$. |
| $P_k^{max,a}$ | Normal capacity (i.e., rate A) for the corresponding transmission asset (thermal limit or stability limit). Typically, $P_k^{min,a} = -P_k^{max,a}$. |
| $P_k^{max,c}$ | Emergency capacity (i.e., rate C) for the corresponding transmission asset. Typically, $P_k^{min,c} = -P_k^{max,c}$. |
| $P_n^{max}$ | Real power maximum capacity for the generator at node $n$. |
| $PTDF_{k,n}^R$ | Proportion of flow on transmission asset $k$ resulting from injection of one MW at node $n$ and a corresponding withdrawal of one MW at reference node $R$. |
| $\bar{\gamma}_{n'(c),n}$ | Recognizes the node with generator loss under contingency $c$ (1 if contingency node, else 0). |

*Variables*

| | |
|---|---|
| $D_n$ | Real power demand at bus $n$ (assumed to be perfectly inelastic). |
| $F_k^-, F_k^+$ | Pre-contingency flowgate marginal prices. Dual variables (or shadow prices) on transmission asset $k$'s normal capacity constraints; lower and upper bounds respectively. |
| $F_k^{c-}, F_k^{c+}$ | Post-contingency flowgate marginal prices. Dual variables on critical transmission asset $k$'s emergency capacity constraints under critical contingency $c$; lower and upper bounds respectively. |
| $P_n$ | Real power production from the generator at node $n$. |
| $\alpha_n$ | Dual variable on generator's (at node $n$) capacity constraint; upper bound constraint. |
| $\delta$ | Dual variable on system-wide power balance constraint (marginal energy component of LMP). |
| $\lambda_n$ | Locational marginal price (LMP) at node $n$. Dual variable that signifies the increase (or decrease) to the primal objective if there is slightly more (or less) consumption by the demand at node $n$. |

## II. Introduction

ISOs maintain a continuous, reliable, and economically efficient supply of electric energy with the assistance of energy management systems and market management systems (MMSs). One key feature within the MMSs is the determination

This work was supported by the Consortium for Electric Reliability Technology Solutions (CERTS) with the US Department of Energy (DOE), and the Advanced Research Projects Agency – Energy (ARPA-E) with the US DOE.

N. G. Singhal is with the Electric Power Research Institute, Palo Alto, CA 94304 USA (e-mail: nsinghal@epri.com). J. Kwon is with Argonne National Laboratory, Lemont, IL 60439 USA (e-mail: kwonj@anl.gov). K. W. Hedman is with the School of Electrical, Computer, and Energy Engineering, Arizona State University, Tempe, AZ 85287 USA (e-mail: khedman@asu.edu).

of the generation dispatch and ancillary services schedule while respecting complex operational requirements and strict physical restrictions. The transmission planning standard (TPL-001-4), set by the NERC is an instance of one such requirement, which stipulates system performance requirements under both normal and emergency conditions [1]. Particularly, the system is required to recover from the loss of any single bulk element, e.g., a generator or a non-radial transmission element, without inconveniencing customers (involuntary load shedding). This rule is more commonly referred to as the *N*-1 reliability requirement and makes the underlying problem stochastic in nature. However, modeling such uncertain events within resource scheduling tools presents two practical barriers: (1) computational complexity of the resulting stochastic optimization problem, and (2) market barriers primarily due to the complications associated with pricing in a stochastic market environment. Consequently, most of the contemporary power system operational frameworks rely on deterministic approaches and utilize numerous approximations to handle uncertainties to meet the *N*-1 mandate.

Today, ISOs model critical transmission contingencies in the market explicitly without utilizing second-stage recourse decision variables; post-contingency line flows are represented using shift factors, such as LODFs, for a subset of critical transmission contingencies. Decomposition techniques are leveraged to manage the complexity of the overall mathematical program by acknowledging only the constraints deemed to be critical. Such approaches enable an efficient handling of critical transmission contingencies within MMSs. A generator outage can also constrain the transmission system considerably. Generator contingencies are not modeled explicitly within state-of-the-art market auction models; instead, system or zonal operating reserve requirements are formed to ensure the system is reliable against generator contingencies. For instance, common industry practices, to approximate the *N*-1 mandate for generator contingencies, include simplistic policies that require a MW level of contingency reserve to be acquired somewhere in the system [2], [3]. However, such policies do not ensure reliable operations (or reserve deliverability) since they only capture a quantitative aspect [4], [5]. Moreover, such approximate, deterministic approaches require out-of-market corrections (OMCs) to adjust resource schedules to account for modeling inaccuracies [6], [7]. Consequently, there is a push in the industry to include an explicit representation of generator contingencies in the market auction models.

Two-stage scenario-based stochastic programs are often proposed to improve operations by optimizing the system response, e.g., reserve activation, in the post-contingency states. However, recent industry movement to model generator contingencies suggests using pre-determined factors, such as generator loss distribution factors (GDFs) and zonal reserve deployment factors [8], [9], to approximate the system response to a generator contingency; such factors are analogous to the more familiar participation factors that are used today in real-time contingency analysis when simulating generator contingencies. CAISO recently proposed to update its market auction models to recognize the impact of generator contingencies and remedial action schemes (RAS) in the market, explicitly, without using second-stage recourse decision variables [8]. Furthermore, MISO augmented their market auction models by modeling the loss of the largest generator for each zone and the corresponding system response in the post-contingency state, explicitly, without using second-stage recourse decision variables [9]. MISO's approach approximates post-contingency congestion on critical transmission interfaces due to the deployed zonal reserves. Moreover, the system response is modeled via zonal aggregated sensitivity factors and pre-determined zonal reserve deployment factors. With the explicit modeling of generator contingencies within the market auction models, the industry is moving away from deterministic program formulations to a stochastic program structure. The anticipated impacts include market prices that better reflect the quality of service provided by generators in response to a generator contingency. The main purpose of this paper is to provide a theoretical analysis of the recent changes in market auction models while focusing on its influences on market clearing prices, i.e., locational marginal prices (LMPs). It investigates the impact that the explicit inclusion of generator contingencies will have on the market pricing structure using duality theory. Primal and dual formulations of market auction models, with and without explicit generator contingency modeling, derivation of the corresponding LMPs to demonstrate how the proposed changes affect market prices, settlements and revenues, are presented.

The remainder of the paper is organized as follows. Section III introduces a theoretical analysis of a contemporary market auction model. Section IV investigates the anticipated changes by providing an enhanced primal formulation for the market auction model and an economic interpretation of the corresponding dual problem, its variables, and its constraints. One goal is to investigate CAISO's newly proposed (and related) payment structure in greater detail. Note that, in the following discussions, GDFs are used to model the corrective actions approximately without using a recourse decision variable. Finally, Section V concludes this paper and summarizes potential future work.

### III. DUAL PROBLEMS OF ELECTRIC ENERGY MARKET FORMULATIONS

#### A. Background on Duality Theory for Linear Optimization

In linear optimization, there is the primal problem, the problem at hand. In this case, the problem of interest is the direct current optimal power flow (DCOPF) problem or a security-constrained economic dispatch (SCED) problem. Each primal problem then has a corresponding dual problem and together they form what is known in linear optimization theory as a primal-dual pair. The dual problem can be interpreted as an optimization problem that is searching for the tightest lower bound (when the primal is a minimization problem); it also provides the shadow prices (dual variables) corresponding to the constraints in the primal. Dual variables, based on linear optimization theory, can also be interpreted as the corresponding Lagrange multipliers for the constraints within the primal. From the perspective of an economist, they are interpreted as shadow prices. The constraints within the dual describe the relationships between the dual variables. Likewise, the primal variables are the corresponding shadow prices for the dual constraints.

The following example demonstrates the relationship between the primal-dual pair, where $\mathbf{a_i}$ is a row and $\mathbf{A_j}$ is a column from a given $\mathbf{A}$ matrix that captures the constraint set (which consists of $M$ constraints: $M_1 \geq$ constraints, $M_2 \leq$ constraints, and $M_3 =$ constraints) for the primal; each constraint has a scalar $b_i$. In addition, $\mathbf{c}$ is the cost vector and $\mathbf{X}$ is the vector of primal variables, where $N_1$, $N_2$, and $N_3$ denote the subset of non-negative, non-positive and unrestricted primal variables respectively. Also, $\mathbf{p}$ denotes the penalty (or shadow) price for violating the corresponding primal constraint. This primal-dual pair presentation can be found in a variety of textbooks, including [10].

Primal:
Minimize: $\mathbf{c^T X}$
Subject to:
$\mathbf{a_i^T X} \geq b_i, \quad i \in M_1$
$\mathbf{a_i^T X} \leq b_i, \quad i \in M_2$
$\mathbf{a_i^T X} = b_i, \quad i \in M_3$
$x_j \geq 0, \quad j \in N_1$
$x_j \leq 0, \quad j \in N_2$
$x_j:$ free, $\quad j \in N_3$

Dual:
Maximize: $\mathbf{p^T b}$
Subject to:
$p_i \geq 0, \quad i \in M_1$
$p_i \leq 0, \quad i \in M_2$
$p_i:$ free, $\quad i \in M_3$
$\mathbf{p^T A_j} \leq c_j, \quad j \in N_1$
$\mathbf{p^T A_j} \geq c_j, \quad j \in N_2$
$\mathbf{p^T A_j} = c_j, \quad j \in N_3$

Prior work related to optimization problems for power systems derive the properties of the prices, which come from the dual formulation, based on applying Karush-Kuhn-Tucker conditions and simplifying the equations [11]-[20]. Note that the dual formulation is derived by creating a Lagrangian dual and then simplifying it into the form presented above. Leveraging the known properties for a primal-dual pair for linear optimization models is used in this paper since it is more concise and straightforward [14].

*B. The Dual Formulation for a Standard DCOPF Problem*

This subsection provides an explicit formulation of the primal-dual pair for the DCOPF problem, which is a simplified representation of existing market formulations that generally come in the form of a security-constrained unit commitment (SCUC) or a SCED model. Most of the contemporary market models use a linearized DCOPF formulation that is based on PTDFs instead of the *B-θ* formulation that relies on the susceptance of transmission assets (*B*) and the bus voltage angles (*θ*). The PTDF-based formulation is easier to solve since it provides the option of ignoring the transmission assets that are inconsequential (i.e., *rarely* congested), thereby reducing modeling complexity. A primal problem formulation for a standard PTDF-based DCOPF is detailed below.

$$\underset{P_n, D_n}{\text{Minimize:}} \sum_n c_n P_n \tag{1}$$

Subject to:
$$-P_n \geq -P_n^{max}, \forall n \in N \quad (\alpha_n) \tag{2}$$
$$\sum_n PTDF_{k,n}^R (P_n - D_n) \geq -P_k^{max,a}, \forall k \in K \quad (F_k^-) \tag{3}$$
$$-\sum_n PTDF_{k,n}^R (P_n - D_n) \geq -P_k^{max,a}, \forall k \in K \quad (F_k^+) \tag{4}$$
$$\sum_n P_n - D_n = 0 \quad (\delta) \tag{5}$$
$$D_n = \overline{D_n}, \forall n \in N \quad (\lambda_n) \tag{6}$$
$$P_n \geq 0.$$

The objective, (1), is to minimize the linear operating costs, which is equivalent to maximizing the market surplus since the demand is assumed to be perfectly inelastic. Constraint (2) imposes an upper bound on the real power scheduled from a generating resource. Note that, for simplicity, the minimum real power generating capacity is assumed to be zero for all generating resources. The dc power flow on a transmission asset is constrained by its normal (thermal or stability) rating, i.e., rate A, in (3) and (4) respectively. The dual variables of (3) and (4) are the flowgate marginal prices for those transmission assets or flowgates; these dual variables are used to calculate the congestion component of the LMP. Constraint (5) assures system-wide power balance between generation and demand; the dual variable of (5) captures the energy component of the LMP. Note that, in this formulation, the demand is treated as a variable following which it is fixed to equal a parameter in (6). The dual variable of (6) signifies the increase (or decrease) to the primal objective (1) if there is slightly more (or less) consumption by the demand at node *n*, which directly translates into the definition of the LMP. The corresponding dual problem formulation for the primal problem is given below.

$$\underset{\alpha_n, F_k^-, F_k^+, \delta, \lambda_n}{\text{Maximize}} : -\sum_n P_n^{max} \alpha_n - \sum_k P_k^{max,a}(F_k^- + F_k^+) + \sum_n \overline{D_n} \lambda_n \tag{7}$$

Subject to:
$$-\alpha_n + \sum_k PTDF_{k,n}^R (F_k^- - F_k^+) + \delta \leq c_n, \forall n \in N \quad (P_n) \tag{8}$$
$$\sum_k PTDF_{k,n}^R (F_k^+ - F_k^-) - \delta + \lambda_n = 0, \forall n \in N \quad (D_n) \tag{9}$$
$$\alpha_n \geq 0, F_k^- \geq 0, F_k^+ \geq 0, \delta \text{ free}, \lambda_n \text{ free}.$$

At optimality, the dual objective (7) is equal to the primal objective (1) by strong duality. The first, second, and the third components of (7) denote the generation rent (short-term generation profit), the congestion rent, and the load payment, respectively. Since generation revenue is equal to generation cost plus generation rent, it can be proven that, at optimality, load payment is equal to generation revenue plus congestion rent. Constraints (8) and (9) represent the dual constraints corresponding to the generator production and the demand variables in the primal problem, respectively. Constraint (9) within the dual problem identifies $\lambda_n$ as the LMP at node *n*. Thus, the LMP, defined by (9a), is equal to the sum of the marginal energy ($\delta$) and the marginal congestion components. Note that, since the PTDF-based DCOPF formulation defined by (1)-(6) is assumed to be a lossless model, there is no loss component of the LMP for the work presented in this paper. After identifying the equation that defines the LMP via (9), (8) reduces to (8a).

$$\lambda_n = \delta + \sum_k PTDF_{k,n}^R (F_k^- - F_k^+), \forall n \in N \quad (D_n) \tag{9a}$$
$$-\alpha_n + \lambda_n \leq c_n, \forall n \in N. \quad (P_n) \tag{8a}$$

The dual variable of (2), i.e., non-negative, signifies the marginal value of increasing a specific generator's maximum capacity. Three cases can potentially exist in this context, while remembering that the lower bounds of all generators are assumed to be zero for this simplified DCOPF problem: 1) if a generator is producing, but not at its maximum capacity (i.e., $\alpha_n = 0$ by complementary slackness), then the LMP at the node of the generator is equal to its marginal cost (by complementary slackness); 2) if a generator is not producing anything (i.e., $\alpha_n = 0$ by complementary slackness), then the LMP at its node is less than or equal to its marginal cost; and 3) if the generator is producing at its maximum capacity (i.e., $\alpha_n \geq 0$), then the LMP at its node is greater than (when $\alpha_n > 0$) or equal (when $\alpha_n = 0$) to its marginal cost (by complementary slackness). These results match with a simple economic interpretation of the shadow price for (2). Whenever you are not producing at your maximum capacity the short-term marginal benefit to increase your capacity beyond its existing capability is zero. When you are operating at your maximum capacity, the short-term marginal benefit to increase your capacity by 1 MW is equal to the difference between your LMP and your marginal cost; keep in mind that this explanation applies to the presented primal DCOPF formulation. If the presented formulation included other constraints that could restrict the generator's output, e.g., ramp rate limits, then the description would be more complex. In general, note that if the DCOPF is formulated differently, the dual will not be the same and may result in different interpretations of that different dual.

## IV. RECENT INDUSTRY MOVEMENTS TO MODEL GENERATOR CONTINGENCIES IN MARKET

### A. Primal Formulation of the Enhanced DCOPF Problem

To meet NERC's *N*-1 mandate more appropriately, recent literature suggests enhancing generator contingency modeling by ensuring post-contingency transmission security through an explicit representation of post-contingency congestion patterns for critical generator contingencies within the market auction models [4], [5], [8], [9]. The mathematical (primal) auction formulation for the enhanced DCOPF problem, motivated by the optimization problem proposed by CAISO in [8], is detailed below. Extensions to this work can be made to analyze other attempts to introduce more advanced corrective control actions.

$$\underset{P_n, D_n}{\text{Minimize}}: \sum_n c_n P_n \qquad (10)$$

Subject to:
$$-P_n \geq -P_n^{max}, \forall n \in N \qquad (\boldsymbol{\alpha_n}) \quad (11)$$
$$\sum_n PTDF_{k,n}^R (P_n - D_n) \geq -P_k^{max,a}, \forall k \in K \qquad (\boldsymbol{F_k^-}) \quad (12)$$
$$-\sum_n PTDF_{k,n}^R (P_n - D_n) \geq -P_k^{max,a}, \forall k \in K \qquad (\boldsymbol{F_k^+}) \quad (13)$$
$$\sum_n PTDF_{k,n}^R (P_n + GDF_{n'(c),n} P_{n'(c)} - D_n) \geq -P_k^{max,c}, \forall k \in K^{crt}, c \in C^{g^{crt}} \qquad (\boldsymbol{F_k^{c-}}) \quad (14)$$
$$-\sum_n PTDF_{k,n}^R (P_n + GDF_{n'(c),n} P_{n'(c)} - D_n) \geq -P_k^{max,c}, \forall k \in K^{crt}, c \in C^{g^{crt}} \qquad (\boldsymbol{F_k^{c+}}) \quad (15)$$
$$\sum_n P_n - D_n = 0, \qquad (\boldsymbol{\delta}) \quad (16)$$
$$D_n = \overline{D_n}, \forall n \in N \qquad (\boldsymbol{\lambda_n}) \quad (17)$$
$$P_n \geq 0.$$

where:
$$GDF_{n'(c),n} = \begin{cases} -1, & n = n'(c) \\ 0, & n \neq n'(c) \land n \notin S^{FR} \\ \dfrac{u_n P_n^{max}}{\sum_{\substack{n \in S^{FR} \\ n \neq n'(c)}} u_n P_n^{max}}, & n \neq n'(c) \land n \in S^{FR} \end{cases}$$
$$\forall n \in N, c \in C^{g^{crt}}. \qquad (18)$$

In this formulation, the generation loss is distributed across the system via GDFs and is presumed to be lossless. Also, it is prorated based on the maximum online (frequency responsive) capacity, to approximate the actual system behavior, while ignoring capacity and ramp rate restrictions [8]. Equation (18) provides CAISO's definition for GDFs, which is proposed to estimate the effect of generation loss and the associated system response on critical transmission assets in the post-contingency state. The post-contingency dc power flow (under a critical generator outage) on a critical transmission asset is restricted by its emergency rating (rate C) in (14) and (15) respectively. The remainder of the formulation is consistent with the standard DCOPF formulation. Note that GDF is constructed to denote the outage of generation at a specific node, not to distinguish between the outage of a single generator at a node with multiple generators. Lastly, CAISO's actual market model will be more complex than what is presented in (10)-(18), which does not include other modeling issues like transmission contingency modeling, reserve requirements, ramp rate limits, etc. The formulation is kept in a simpler manner to focus on the key proposed change, which is related to the inclusion of the generator contingency modeling with the use of the GDFs.

### B. Dual Formulation for the Enhanced DCOPF Problem

The corresponding dual problem formulation is described below. Note that, while the following dual is derived based on the formulation in Section IV.A, other primal formulations are also possible, and the dual formulations will change as well.

$$\underset{\alpha_n, F_k^-, F_k^+, F_k^{c-}, F_k^{c+}, \delta, \lambda_n}{\text{Maximize}}: -\sum_n (P_n^{max} \alpha_n) - \sum_k \left(P_k^{max,a}(F_k^- + F_k^+)\right)$$
$$-\sum_{\substack{k \in K^{crt} \\ c \in C^{g^{crt}}}} (P_k^{max,c}(F_k^{c-} + F_k^{c+})) + \sum_n (\overline{D_n} \lambda_n) \qquad (19)$$

Subject to:
$$-\alpha_n + \sum_k PTDF_{k,n}^R (F_k^- - F_k^+) + \left(\sum_{\substack{k \in K^{crt} \\ c \in C^{g^{crt}}}} (F_k^{c-} - F_k^{c+}) \left(PTDF_{k,n}^R + \bar{\gamma}_{n'(c),n} \sum_{s \in N} PTDF_{k,s}^R GDF_{n'(c),s}\right)\right) + \delta \leq c_n, \forall n \in N \qquad (\boldsymbol{P_n}) \quad (20)$$
$$\sum_k PTDF_{k,n}^R (F_K^+ - F_k^-) + \sum_{\substack{k \in K^{crt} \\ c \in C^{g^{crt}}}} PTDF_{k,n}^R (F_K^{c+} - F_k^{c-}) - \delta + \lambda_n = 0, \forall n \in N \qquad (\boldsymbol{D_n}) \quad (21)$$
$$\alpha_n \geq 0, F_k^- \geq 0, F_k^+ \geq 0, F_k^{c-} \geq 0, F_k^{c+} \geq 0, \delta \text{ free}, \lambda_n \text{ free}.$$

where:
$$\bar{\gamma}_{n'(c),n} = \begin{cases} 0, & n \neq n'(c) \\ 1, & n = n'(c) \end{cases}, \forall n \in N, c \in C^{g^{crt}} \qquad (22)$$

$$\lambda_n = \delta + \sum_k PTDF_{k,n}^R(F_k^- - F_k^+) + \sum_{\substack{k \in K^{crt}, \\ c \in C^{g^{crt}}}} PTDF_{k,n}^R(F_k^{c-} - F_k^{c+}), \forall n \in N. \quad (D_n) \quad (21a)$$

The dual objective now has an additional term, i.e., the third component in (19), which represents the post-contingency congestion rent resulting from generator contingency modeling. Constraints (20) and (21) represent the dual constraints corresponding to the generator production and the demand variables in the enhanced primal problem respectively. The primary impact that the proposed changes will have on market pricing is how it affects the LMPs. Constraint (21) within the dual problem identifies $\lambda_n$ as the LMP at node $n$, which is further defined in (21a) and is equal to the sum of the marginal energy component, the marginal pre-contingency congestion component, and an additional marginal post-contingency congestion component that comes from the modeling of critical generator contingencies.

Note that, the enhanced primal problem defined by (10)-(18) differs from CAISO's primal problem in [8] with respect to the following aspects. 1) The demand is first treated as a variable following which it is fixed to equal a parameter in (17) to enable the derivation of the LMP in a simpler manner. 2) The losses are ignored for simplification. 3) No new variables are introduced to denote the real power production from the system's resources under a specific generator contingency. 4) Transmission contingency security constraints are ignored to allow the derivation in this paper to focus on generator contingency modeling in a clear and concise manner.

*C. Analyzing the Dual Formulation*

To understand what is communication by the aforesaid dual problem, first, start with the objective functions of the primal and the dual. Linear optimization theory includes strong duality, which guarantees that the objective of the primal problem equals the objective of the dual problem at optimality. Achieving strong duality means there is no duality gap. Another way to interpret the strong duality relationship is through its expression of an exchange of money; payments and expenses resulting from the auction and the corresponding exchange of goods and services. There is the obvious piece from the objective of the primal problem, which is the total generation cost. The next obvious piece is the last term of the dual objective, which is the load payment, LMP times consumption. The first term of the dual objective is the short-term generator profit for a generator at node $n$, summed over all nodes, or the system-wide generation rent. The second and third terms in the dual's objective represent the system-wide congestion rent; this is to be expected as it relates the flowgate marginal price to the line flow, once complementary slackness is applied.

The discussion that follows analyzes the corresponding impact on the rent and the revenue for generators that are and are not included in the assumed set of critical generator contingencies. Recall first that GDF reflects the anticipated system response from a specific node in the network and the way it is structured in (18) assumes that there is only one unit at most at a node (easily modifiable). Second, note the formula to define the GDF is based purely on the generator's capacity relative to the rest of the fleet's capacity (for units that are frequency responsive). One obvious drawback of the proposed GDF is that it ignores the generator's capacity, the generator's ramp rate restrictions, and whether the ISO procured the necessary reserve product from the unit. As a result, this model assumes there is the capability to inject power at a node based on the definition of the GDF, not necessarily based on the actual ability for the generator to provide the needed reserve; for instance, a generator may be operating at its maximum capacity. The assumed GDF only accounts for capacity while not capturing the dispatch set point of the unit or whether the unit has been obligated to provide contingency reserve products (note that while the presented auction formulation does not include reserve procurement, the GDF itself does not reflect whatsoever on reserve and, as such, the impact of reserve is not captured anyway). Finally, the GDF shows up only in (14) and (15) and it is multiplied by the MW dispatch variable for the generator that is modeled to be under outage (or contingency). This translates into the post-contingency congestion for a generator outage, in the primal problem, to being directly related to the dispatch variable for the contingency generator only. For example, assume that generator 1 is lost, which is located at bus 1. Assume that generator 2 is anticipated to completely pick up this entire loss of supply from generator 1 (so the ISO sets the GDF = 1 for generator 2) and generator 2 is at bus 2. Even though generator 2 is the unit anticipated to provide the needed injection, the functional form in (14) and (15) relate the change in the injection at bus 2 to be determined by the GDF (a fixed input parameter) and the output of generator 1. The GDF is basically masking the response that is provided by generator 2 for generator 1's drop in supply. Thus, $\bar{\gamma}$ reduces to zero in (20) for generators that are not located at the nodes of the generators contained within the critical generator contingency list.

More importantly, the post-contingency congestion and the component of the LMPs that are reflective of this post-contingency state are driven by the cost of the generator that is lost; it is not driven by the cost that is associated to the generators that would respond. From a power engineering perspective, capturing the costs of the units responding does not fully matter in regard to ensuring a secure system; the primal problem captures the change in injection for the buses that are anticipated to have an increase in production when responding to the outage. From a cost perspective, it has an impact as the cost is only related to the generator that is lost and not the units that respond. For instance, suppose that the only generator contingency that is explicitly modeled is a large, baseload unit like a nuclear unit. Such baseload units are often cheaper in regard to their marginal cost, $/MWh. On the other hand, the units that are likely to respond will be units that have fast ramping capabilities and are flexible; those are units that are generally more expensive. With this example it is clear that, at the very least, there is a high probability that the units that are chosen to be in the critical contingency list may be rather distinct in characteristics, and costs, than the units that are expected to respond. This is important since, again, the cost in

the post-contingency state is not driven by the units responding but by the unit that is lost. At the same time, the model does not acknowledge any costs due to re-dispatch of generators in the post-contingency state. Only the pre-contingency costs are considered, which makes the impact of the post-contingency congestion a secondary influencing factor; the cost changes only by forcing a different pre-contingency dispatch set point that is secure rather than acknowledging the change in dispatch cost due to activation of reserve. Last, it is also equally important, if not more, to acknowledge the influence this has on the prices, via duality theory, that are then produced by this proposed reformulation by CAISO. This paper does not dive deeper into this potential problematic issue since that is a topic that will require much more research and investigation, as identified in the future work in Section V.

For generators that are not included in the critical list, using the definition of $\bar{\gamma}$ from (22) and the LMP for that generator's location ($\lambda_n$) from (21a), (20) can be rewritten as (20a). Thus, (8a) and (20a) are similar with the exception that the LMP is now capturing an added, new congestion component reflecting congestion in the post-contingency operational state with the loss of a generator. Complementary slackness is then applied to the constraint-dual variable pair in (20a) to create (23). It is noteworthy to emphasize that (23) turns out to be identical to what would be obtained by applying complementary slackness to (8a), which again allows for the determination of the generator rent. Complementary slackness can also be applied to (11) to form (24). Then, based on (23) and (24), the generator rent for a generator that is not in the assumed critical generator contingency list is given by (25). The short-term generator profit (or generator rent) that will be earned by the non-critical generators is equal to the generator revenue less the generator cost. This generator rent term is basically identical to what is seen from the standard DCOPF formulation excepting that the LMP has an additional congestion component.

$$-\alpha_n + \lambda_n \leq c_n \qquad (P_n) \qquad (20a)$$
$$-\alpha_n P_n + \lambda_n P_n = c_n P_n \qquad (23)$$
$$-P_n \alpha_n = -P_n^{max} \alpha_n, \forall n \in N \qquad (24)$$
$$P_n^{max} \alpha_n = \lambda_n P_n - c_n P_n. \qquad (25)$$

For the critical generators, using the definition of $\bar{\gamma}$ from (22) and the LMP for that generator's location ($\lambda_n$) from (21a), (20) can be rewritten as (20b). Complementary slackness is then applied to the constraint-dual variable pair in (20b) to create (26). Then based on (26) and (23), the generator rent for a generator that is contained within the assumed critical generator contingency list is given by (27).

It is pertinent to note that the LMP defined in (21a) is consistent with CAISO's LMP definition in [8] for the nodes that do not have critical generators. Furthermore, CAISO's LMP definition for the nodes that do have critical generators, i.e., whose outages are modeled explicitly, is provided below in (21b). Note that there is a resemblance in the last term of the profit function (within square brackets) for the generators that are included in the critical contingency list in (27) and the last term of CAISO's LMP definition in (21b). This entails a detailed investigation into the net revenue stream for the generators that are contained within the assumed critical generator contingency list.

$$-\alpha_n + \lambda_n + \sum_{\substack{k \in K^{crt}, \\ c \in C^{g^{crt}}}} [(F_k^{c-} - F_k^{c+})(\sum_{s \in N} PTDF_{k,s}^R GDF_{n'(c),s})] \leq c_n \qquad (P_n) \qquad (20b)$$

$$-\alpha_n P_n + \lambda_n P_n + \sum_{\substack{k \in K^{crt}, \\ c \in C^{g^{crt}}}} [(F_k^{c-} - F_k^{c+})(\sum_{s \in N} PTDF_{k,s}^R GDF_{n'(c),s})] P_n = c_n P_n \qquad (26)$$

$$P_n^{max} \alpha_n = \lambda_n P_n - c_n P_n + \sum_{\substack{k \in K^{crt}, \\ c \in C^{g^{crt}}}} [(F_k^{c-} - F_k^{c+})(\sum_{s \in N} PTDF_{k,s}^R GDF_{n'(c),s} P_s)]. \qquad (27)$$

CAISO's LMP definition for nodes with critical generators [8]:
$$\lambda_n = \delta + \sum_k PTDF_{k,n}^R (F_k^- - F_k^+) + \sum_{\substack{k \in K^{crt}, \\ c \in C^{g^{crt}}}} PTDF_{k,n}^R (F_k^{c-} - F_k^{c+}) + \sum_{\substack{k \in K^{crt}, \\ c \in C^{g^{crt}}}} [(F_k^{c-} - F_k^{c+})(\sum_{s \in N} PTDF_{k,s}^R GDF_{n'(c),s})] \qquad (21b)$$

There are a few key issues to consider here. First, note that despite the presence of the extra term in (27); (27) describes the profit that will be earned by the critical generators. In this specific primal reformulation of the DCOPF, described by (10)-(18), only a single linear operating cost component is considered for the production from the generator at node $n$. In addition, the generator production variable has no other restrictions other than lower and upper bound restrictions. Analogous to the dual analysis provided for the standard DCOPF problem in Section III.B, three cases can potentially exist for the generators in this primal reformulation with single linear cost coefficients and only lower and upper bounds. (1) If a generator is not producing at its maximum capacity, the short-term marginal benefit (profit) to increase its capacity beyond its existing capability is zero. (2) If it is operating at its maximum capacity, the short-term marginal benefit to increase its capacity by 1 MW is equal to the difference between what it is paid and its marginal cost. (3) If it is not producing anything, then what it is paid must be less than or equal to its marginal cost. Consequently, the penalty (shadow) price of (11), $\alpha$, does completely capture the $/MWh rate for the corresponding unit's profit or the marginal benefit of increasing its maximum capacity, which makes $\alpha_n P_n$ (or $\alpha_n P_n^{max}$ by complementary slackness) its profit function. Therefore, the net revenue stream for a generator should equal the sum of its profit and cost. Equation (27a) defines the revenue for the generators that are contained within the assumed critical list of generators.

$$P_n^{max} \alpha_n + c_n P_n = \lambda_n P_n + \sum_{\substack{k \in K^{crt}, \\ c \in C^{g^{crt}}}} [(F_k^{c-} - F_k^{c+})(\sum_{s \in N} PTDF_{k,s}^R GDF_{n'(c),s} P_s)]. \qquad (27a)$$

Second, if the short-term generator profit for a generator at node $n$, summed over all nodes, is equal to the total generation profit, then by strong duality, at optimality (28) holds. In other words, at optimality, the objective of the primal reformulation must equal the objective of the dual problem, or the load payment is equal to the sum of the total generation cost, the total generation profit, and the total congestion rent.

$$\sum_n c_n P_n = -\sum_n P_n^{max} \alpha_n - \sum_k P_k^{max,a}(F_k^- + F_k^+) - \sum_{\substack{k \in K^{crt}, \\ c \in Cg^{crt}}} P_k^{max,c}(F_k^{c-} + F_k^{c+}) + \sum_n \overline{D_n} \lambda_n. \quad (28)$$

If $\alpha_n P_n^{max}$ is equal to the short-term generator profit for a generator then it implies that (27) describes the profit that will be earned by the critical generators, which means that a critical unit's revenue is not just the LMP at its location ($\lambda_n$) times the production, but its revenue also includes the added extra (last) term in (27). On the contrary, if $\alpha_n P_n^{max}$ is not equal to the short-term generator profit and instead a critical unit is only paid the LMP at its location ($\lambda_n$) times the production, then its profit is equal to LMP at its location ($\lambda_n$) times the production less the cost. In this case, the short-term generator profit for a generator at node *n*, summed over all nodes, will not equal the term in the dual objective that is supposed to represent the total generation profit of the entire system. In other words, this would remove the added extra term from its revenue. Furthermore, since complementary slackness dictates that (27) should hold, which again is the short-term generator profit for a generator at node *n*, summed over all nodes, will not equal the total generation profit of the entire system. To summarize, if $\alpha_n P_n^{max}$ does *not* denote the short-term generator profit for both the critical and the non-critical generators, it will result in an ISO that is *not* revenue neutral. The ISO will either have a revenue shortfall overall or surplus. This confirms and clarifies what the generators in the assumed critical list should be paid and explains the reasoning behind why CAISO is potentially including the added extra term in their definition of the LMP at the nodes of the critical generators in (21b). However, CAISO's definition of the LMP at the nodes of the critical generators is *not* consistent with the traditional definition of the LMP or the LMP that is identified by the corresponding dual formulation. The critical generators should now be paid this extra term but that does not imply the extra term be included in the LMP since this will have associated implications in the FTR markets (note that this LMP is further used to settle the FTR payments in the FTR markets). Also, note that, in (28), the load pays the LMP identified by the dual formulation. Further extension of this work is necessary to evaluate more advanced reformulations to enhance generator contingency modeling (an example is detailed in Section V) and its corresponding effect on market prices, settlements, and revenues.

Third, it is necessary to understand the interpretation and the implications of the extra term in (27). If a critical generator is under an outage, the GDFs specify that the corresponding injections to compensate for the drop in its supply are defined based on its value for its locations. Essentially, the extra term in the profit function, (27), is what the critical generator is paid by the ISO to compensate for the loss of its production. Now, if the extra term is combined with the fact that the unit is being paid the LMP at its location for its production, what this translates into is that the corresponding critical generator is basically paying a congestion charge for the difference between injecting at its location and instead injecting at the locations identified by the GDFs. Thus, the combination of the extra term and its LMP component corresponding to the outage is basically a congestion transfer cost. Another way to interpret this is that, for that particular contingency scenario, the critical generator will inject at the locations that are identified by the GDFs instead of injecting at its own location. The LMP already compensates for the expected injection at its location. The extra term is the transfer due to injections based on the pre-defined rules of the GDF. Another way to interpret this would be that the generator that is lost has storage at each node identified by the GDF definition and is expected to compensate for its own contingency by injecting at those locations. The model still acknowledges that the generator is producing; it is just producing now magically at different locations. As such, the dual formulation suggests that the unit should be compensated exactly by that (invalid) assumption. The right way to make this work is to have the critical generator buy from the locations identified by the GDF instead or have some sort of a side contract with the generators at those locations. This paper defines how this pricing structure would work if it is to follow the exact prescribed formulation proposed within the primal.

To assist in understanding the implications, it is helpful to go back to (14) and (15). Recall that the GDF shows up only in (14) and (15) and it is multiplied by the MW dispatch variable for the generator that is lost. This basically simulates the cost of a critical generator backing up its loss based on its costs at other locations. The fact that the equations are driven based on its dispatch variable and not the dispatch variables of the responding generators implies that this critical generator's cost influences the duals for this issue and not the cost of the responding generators. Thus, the responding generators do not affect the outcome for the generator that is lost. This is an important implication because this is sensitive to which generators are chosen to be included in the list of critical generator contingencies. CAISO acknowledges this concern by stating that analogous to how transmission security constraints are selectively enforced in contemporary markets, the ISO will decide which generators are critical and need to be explicitly modeled based on engineering analysis and outage studies [8].

Finally, the definition of congestion rent is the flowgate marginal price times the flow on the line, summed over all lines. Congestion rent can be also identified by the difference in load payment and the generation revenue; it is easy to confirm that these two approaches provide the same value for the congestion rent as strong duality provides a formula where generation cost (the primal objective) is equal to the load payment minus the generation rent (short-term generation profit) minus the congestion rent at optimality. This can be more easily identified by applying strong duality to the simplified DCOPF and its dual, (1) and (7). For the more complicated primal reformulation auction model that includes security constraints associated to generator contingencies, it becomes more complex; the congestion rent can be identified by taking (12)-(15) and applying complementary slackness. Thus, the system-wide congestion rent in this case is equal to the second and third terms in the dual's objective, (19); this is to be expected as it relates the flowgate marginal price to the line flow, once complementary slackness is applied.

## V. CONCLUSIONS AND FUTURE RESEARCH

This paper presents a comprehensive theoretical analysis of the recent industry movements, specifically, CAISO's efforts, to model generator contingencies in market models more appropriately. The main intention is to examine (and question) and complement the movement in the industry to enhance generator contingency modeling and to analyze the market impact of the policies that are proposed in this realm of research. It is noteworthy to emphasize that if the primal (DCOPF problem in this context) is formulated differently, the dual will not be the same and may result in different interpretations of that different dual. As this paper has demonstrated, this is why it is very important to perform a rigorous evaluation via duality theory to investigate the potential implications of the proposed market change.

Further extensions of the auction formulation presented in Section IV is essential to evaluate more advanced reformulations to enhance generator contingency modeling and its corresponding effect on market prices, settlements, and revenues. For instance, it is pertinent to analyze an auction reformulation that enhances generator contingency modeling by incorporating an explicit representation of post-contingency power balance in addition to the previously mentioned post-contingency transmission security for critical generator contingencies. The post-contingency power balance constraints will help in assuring system-wide power balance between post-contingency generation and post-contingency demand. This essentially also provides an opportunity to model the expected post-contingency demand consumption under the different critical generator contingency states. However, the obvious setback with such an approach, i.e., the explicit inclusion of post-contingency power balance constraints, is the associated increase in the computational complexity of the corresponding problem. The anticipated impact that the corresponding change will have on market pricing is (again) how it affects the LMPs. This change will result in an LMP at a node for each of the modeled critical generator contingency states. In addition, it is also important to analyze the corresponding effect on the market settlements and revenues.

The explicit consideration of credible generator contingencies in general is expected to result in fewer ex-post OMCs (or adjustments), which is technologically and economically beneficial [4], [5], [7]. The explicit consideration of credible generator contingencies (and fewer ex-post adjustments) enable the market auction to optimize more of the market, which, in turn, results in improvements in market efficiency and improved price signals [4], [5].

Future research should include implementing and testing the effectiveness of the proposed enhancements in improving the market surplus on a large-scale test system. Furthermore, to overcome the issues identified in this chapter, future work should examine the market implications of the generator contingency modeling approach proposed by MISO in [9] and identify a means to extend MISO's approach to include both intra-zonal and inter-zonal transmission assets in addition to modeling more than one critical generator contingencies per reserve zone. The next steps should also investigate new means to introduce corrective actions via different reformulations and the associated market impacts.